\documentclass[aps,pra,preprint,fleqn,groupedaddress,superscriptaddress,amsmath,amssymb,a4paper]{revtex4-1}
\usepackage{graphicx}
\usepackage{color}
\usepackage{ulem}

\newif\ifrev
\revtrue

\begin{document}
	
\title{Radiation Reaction in the Plasma-Based High-Energy Accelerators}
\author{M. R. Islam}
\email{mohammed.islam-3@manchester.ac.uk}
\author{G. Xia}
\email{guoxing.xia@manchester.ac.uk}
\author{Y. Li}
\author{B. Williamson}
\affiliation{School of Physics and Astronomy, University of Manchester, M13 9PL, United Kingdom.}
\affiliation{The Cockcroft Institute, Warrington, WA4 4AD, United Kingdom.}
	

\begin{abstract}
Plasma-based accelerators have achieved tremendous progress in the past few decades, thanks to the advances of high power lasers and the availability of high-energy and relativistic particle beams. However, the electrons (or positrons) accelerated in the plasma wakefields are subject to radiation losses, which generally suppress the final energy gains of the beams. In this paper, radiation reaction in plasma-based high-energy accelerators is investigated using test particle approach. Energy-frontier TeV colliders based on a multiple stage laser-driven plasma wakefield accelerator and a single-staged proton-driven plasma wakefield accelerator are studied in detail. The results show that the higher axial and transverse field gradients seen by an off-axis injected witness beam result in a stronger damping force on the accelerated particles. Proton-driven plasma wakefield accelerated electrons are shown to lose less energy compared to those accelerated in a multi-staged laser-driven plasma wakefield accelerator.  
 \end{abstract}
\maketitle
\section{Introduction}
The plasma-based accelerators have made enormous progress over the last decades. Specifically, the laser-plasma wakefield accelerators (LWFA) are able to produce the shot-to-shot stable and high quality quasi-monoenergetic electron beam up to GeV in centimeter long plasma through self-injection \cite{Mangles2004,Geddes2004,Faure2004,lena+06,Islam15,Brunetti10,clayton+10,pume+02,Esarey2}. With the recent demonstration of a 4.25 GeV beam in a 9 cm long capillary plasma cell \cite{Leemans2} and the successful coupling of two LWFA stages \cite{stti+16}, it is natural to think about the future compact and high energy accelerator based on LWFA scheme \cite{Leemans3}.
However, there are three common factors which limit the energy gain for the LWFA based accelerators. They are phase dephasing, laser depletion and laser diffraction. To be specific, the relativistic witness electrons outrun the plasma wave and get out of the accelerating phase, and stop gaining further energy. Such dephasing can be mitigated by adopting a longitudinal plasma gradient \cite{kat+86, puko+08}. Furthermore, the energy of the laser depletes while driving a plasma wave, which therefore limits the driver's ability to sustain the plasma wave over a longer distance. The third restriction is laser diffraction, which is similar to the expansion of an electron or proton bunch during their evolution. Fortunately, in the case of a discharged capillary, the laser pulse can be guided over longer distances \cite{eskr+94}. Such guiding succeeds when the relativistic self-focusing \cite{ab+92} compensates the diffraction of the laser. 

Currently, intensities up to $10^{22}$ W/cm$^2$ are available in tabletop lasers at their focal spots \cite{yanovsky08} but can be potentially boosted to the order of  $10^{24}$ W/cm$^2$ at the Extreme Light Infrastructure (ELI) \cite{eli}. Electrons interacting directly with lasers in such high intensities can be accelerated to ultra-high energies within a fraction of the laser period and emit a relatively large amount of electromagnetic radiation.  During such interaction, nonlinear effects like radiation reaction due to huge acceleration come into play \cite{jackson99}. Radiation reaction arises from the interaction between the electromagnetic fields and the oscillating electrons. Another important phenomenon likely occurs under high intensities is the production of electron-positron ($e^{-}e^{+}$) pairs from the vacuum and the showers of charged particles \cite{hu16}; and these showers produced by repeated emission of hard photons and pair creation by hard photons. Threshold field strength for such a phenomenon to occur is given by Julian Schwinger \cite{sch51} $E_s = m^2\, c^3/e \hbar \approx1.3 \times10^{18}$ V/m, where, $\hbar$ is the reduced Planck constant, $m$ is the electron mass, $e$ is the electron charge, and $c$ is the speed of light. As to the laser-plasma wakefield accelerator, the transverse field is relatively weaker than the Schwinger field. As oscillation accompanies the electron acceleration, the electrons radiate electromagnetic fields in the forward direction. The radiation spectrum of the ultra-relativistic electrons lies in the X-ray to the gamma-ray range \cite{simo+11}.   In the gamma-ray radiation regime, the integrated radiation loss would be non-negligible compared to the typical energy gain by the electron beam. 

On the other hand, the electron beam driven plasma wakefield accelerators (PWFA) have successfully demonstrated the energy doubling of the 42 GeV electron beam within an 85 cm long plasma at the Stanford Linear Accelerator Center (SLAC) \cite{blcl+07}. The recent experiments have also shown the energetic electron beam acceleration with low energy spread and emittance achieved \cite{Litos} and the positron acceleration in self-loaded plasma wakefield regime \cite{Corde} and in hollow plasma \cite {Gessner}. Therefore it is conceivable that a high energy collider can be designed based on the PWFA scheme \cite{Seryi, Adli}.  However, it has to rely on the multiple stage acceleration to boost the beam energy up to TeV level due to limited energy content in today's electron driver beams.

More recently, high energy proton-driven plasma wakefield acceleration (PDPWA) has been proposed to accelerate an electron bunch up to the energy frontier in a single stage of plasma \cite{caldwell13}. Due to the availability of high energy proton beams and their intrinsic significant energy content, usually two to three orders of magnitude higher than that of the existing energetic electron beams, a single stage PDPWA-based energy frontier collider ($e^{+}e^{-}$ collider or electron-proton ($e$-$p$) collider) can be realised using the existing CERN accelerator complex, e.g. the Super Proton Synchrotron (SPS) or the Large Hadron Collider (LHC) \cite {Guoxing14}. The current AWAKE experiment at CERN will hopefully tackle the fundamental issues concerning the PDPWA-based collider design.

For the aforementioned high energy machines based on plasma accelerators, the energy gain by the charged particle is close to TeV range, and the accelerating beam is inevitably affected by the radiation loss. In this paper, the radiation reaction in the plasma-based accelerators is investigated from the first principle. The energy-frontier TeV colliders based on multiple staged laser-driven plasma wakefield accelerators and a single-staged proton-driven plasma wakefield accelerator are studied in details. The paper is organised as follows.  Section \ref{sec:raddamp} elaborates the theory of radiation reaction. Section \ref{sec:stagelwfa} gives the radiation loss by a test electron accelerating in a laser-driven multi-staged plasma accelerator. Section \ref{sec:protondriv} presents the radiation loss by the charged particles while accelerating in the proton-driven single staged plasma accelerator, with the SPS and the LHC beam as the driver respectively. The conclusions are summarised in Section \ref{sec:conclusion}.

\section{Radiation Damping}
\label{sec:raddamp}
To calculate radiation reaction self-consistently, one should solve the so-called Lorentz-Abraham-Dirac (LAD) equation \cite{jackson99}. This equation has some technical inconsistencies. For instance, the {\it run away} solution in which an electron experiences an exponentially diverging acceleration even in the absence of an external field, violates the causality principle. Typical time scales underneath this methodology are negligible, e.g., for electron, $10^{-23}$s. This time scale is usually insignificant, but recent advances in the high-power and ultra-intense laser technology have raised much interest to look at it \cite{Piazza12}.  Although these problems do not exist in the classical electrodynamics, the radiation can be treated in classical approach if QED parameter $\mathcal{X}= \sqrt{(m c \gamma \vec{E} + \vec{p} \times \vec{H})^2-(\vec{p}\cdot \vec{E})^2}/(mc E_{s}) \simeq \gamma F_{\perp}/(e E_s) << 1$, where $F_{\perp}$ is the transverse force responsible for the oscillation of a charged particle, $\vec{E}$ is the electric field, $\vec{H}$ is the magnetic field, $\vec{p}$ is the momentum, and $\gamma$ is the relativistic Lorentz factor.  
The radiation reaction of a charged particle can be considered as an additional force apart from the Lorentz force. This force damps the particle energy and momentum during the acceleration process, which leads to alteration of the electron trajectory from the predicted one by the individual Lorentz force. According to Landau-Lifschitz \cite{lali80}, considering the radiation reaction force, the equation of motion of the relativistic charged particle with mass $m$ and charge $-e$, in an electromagnetic field, is given by
\begin{equation}
\gamma \frac{d u^{\mu}}{d t} = \frac{c r_e}{e} F^{\mu \nu} u_{\nu} + {2 r^2_e \over 3 mc } F^{\mu}_{rad}\,,
\label{eqn:eqn1}
\end{equation}
where $F^{\mu \nu}$ is the electromagnetic field tensor, $u^{\mu}$ is the 4-velocity of the charged particle, $r_e= e^2/m c^2$, 
$ F^{\mu}_{rad}=(e/r_e)(\partial F^{\mu \nu}/\partial x^{\lambda}) u_{\nu} u^{\lambda} - F^{\mu \lambda} F_{\nu \lambda} u^{\nu} + (F_{\nu \lambda} u^{\lambda})(F^{\nu m} u_m) u^\mu$, the first term in the right hand side of Eq.(\ref{eqn:eqn1}) corresponds to the Lorentz force, and the last term is the radiation reaction force. We normalize  the variables  as $x \rightarrow  k_p x$, $t \rightarrow  \omega_p \, t$, $\vec{v} \rightarrow \vec{v}/c$,  $\vec{p} \rightarrow \vec{p}/(m\, c)$, and $\phi \rightarrow \phi/(m\,c^2)$, where $\omega_p=(4\pi e^2 n_0/m)$ is the plasma frequency, $n_0$ is the background plasma  density, and $k_p=\omega_p/c$. 

When a Gaussian laser pulse propagates in an under-dense plasma, it expels electrons away from the high intensity regions and an ion cavity is created. Similarly, a charged-particle beam can be employed to produce wakefields due to its space charge \cite{essp96,rocl88}. In any plasma-based accelerators, we can model the cavity as a sphere comprising uniform ions with a radius of $r_b \simeq  \pi/k_p $. This value is different in a laser-driven case, where it is determined by the relation $r_b \,k_p\simeq 2 \sqrt{a_0}$, where $a_0 \sim 0.85\times10^{-9}\sqrt{I[W/cm^2]\lambda_0[\mu m]}$ is the normalized laser energy, $I$ is the laser intensity, and $\lambda_0$ is the laser wavelength \cite{luhu+06}. A phenomenological model, where the electric field 
$\vec{E} = [(1+v_p) \xi \hat{x} + \vec{r}_{\perp}]/4$ and magnetic field $\vec{B}=[ z \hat{y} - y \hat{z}]/4 $ are estimated within the bubble, is used by Kostyukov {\it et al.} \cite{kopu+04}. Here $v_p$ is the bubble velocity, $\xi=x- v_p t$ is the longitudinal co-moving coordinate, $x$ is the propagation distance, and $r_{\perp}=\sqrt{y^2+z^2}$. Substituting the modeled  electromagnetic fields into the Eq.({\ref{eqn:eqn1}), we have the equations of motion for a trapped test electron as:
\begin{subequations}
\begin{align}
&\frac{d \xi}{d t} = {{p}_{||} \over \gamma} - v_p \,,\label{eqn:eqn2a}\\
&\frac{d \vec{r}_{\perp}}{d t} = \frac{\vec{p}_{\perp}}{\gamma}\,,\label{eqn:eqn2b}\\
& \frac{d p_{||}}{dt} = -{(1+ v_p) \xi \over 4} + \frac{\vec{r}_{\perp}\cdot \vec{p}_{\perp}}{4\gamma} - {2 e^2 \omega_p \over 3 m c^3} \,\frac{r^2_{\perp} \gamma }{4} p_{||}\,,\label{eqn:eqn2c}\\
& \frac{d\vec{p}_{\perp}}{dt} = -{\gamma+ p_{||} \over \gamma}\, \frac{\vec{r}_{\perp}}{4} - {2 e^2 \omega_p \over 3 m c^3}\,\frac{r^2_{\perp} \gamma }{4}\vec{p}_{\perp}\,,\label{eqn:eqn2d}
\end{align}
\label{eqn:eqn2}
\end{subequations}
where, $p_{||}$ and $p_{\perp}$ are the longitudinal and transverse components of momentum, respectively. Equations (\ref{eqn:eqn2}a-\ref{eqn:eqn2}d) can be numerically solved if the fields are known. The strength of radiation reaction of the electron depends on the electromagnetic fields, the transverse amplitude of oscillations, and the electron energy. For a specific plasma and laser setup, the larger the initial transverse field amplitude is, the greater the radiation damping would be. 

  \section{Multi-staged Laser Wakefield Accelerator}
  \label{sec:stagelwfa}
The multi-staged laser-driven plasma wakefield accelerator has been proposed as a promising approach to achieve a TeV energy gain \cite{lees+09}. Recently at the BELLA centre, scientists have demonstrated that a laser pulse can accelerate an electron beam and then couple it to a second laser plasma accelerator to further amplify its energy gain \cite{stti+16}. In LWFA, the energy gain is estimated based on the plasma density and driver strength. For a specific plasma density, the field strength is defined as $E_x=96\times\sqrt{n_0[cm^{-3}]}$ V/m, and the associated dephasing length is $L_d$=$\lambda_p\gamma^2_g$, where $\gamma_g$=$\omega_0/\omega_p$ is the Lorentz factor defined with respect to laser frequency, $\omega_0$, and $\lambda_p$=$2\pi/k_p$. For a plasma density of $n_0$=$1.4\times10^{17}$cm$^{-3}$, the corresponding field strength, $E_x$=$33$ GV/m and the dephasing length is close to a meter. Therefore, according to the proposed method, one needs more than 30 stages to create a TeV collider. Such gain in energy is subject to the radiation loss. 
\begin{figure}[ht]
     \scalebox{0.65}
     {
          \includegraphics{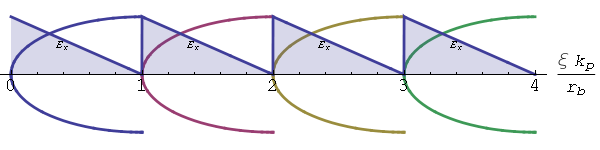} 
      }
 \caption{A schematic diagram of a staged-LWFA with the accelerating phases within the bubbles. }
    \label{fig:scheme}
\end{figure} 

In the experimental setup of this multi-staged accelerator, the plasma cells are interconnected by vacuum region. Each plasma cell is excited by an individual laser pulse at a constant time gap.  At the onset of the process, the first laser excites the plasma wave and generates 33 GeV electron beam in a plasma cell of one dephasing length ($L_d$=1 meter). Once the electron beam reaches the dephasing point, the second plasma cell is excited by a new laser pulse and let the electron beam to place correctly at the accelerating phase. This way one can multiply the energy gain. A field structure is modelled to investigate the radiation loss by the accelerated charged particle in this multi-staged accelerator. The field strength is modelled as $ \vec{E} = (\eta \hat{x} + \vec{r}_{\perp})/4$, where, $\eta=-r_b\lbrace\pi/2 + \arctan(\cot (\pi \xi/r_b))\rbrace/\pi$. The schematic diagram for the bubble structures and the corresponding field strengths are shown in Fig.~({\ref{fig:scheme}). In this model, the gap between the plasma cell and charge-coupling efficiency are neglected.

We solve the Eqs.(\ref{eqn:eqn2}) to evaluate the energy gain and the corresponding radiation loss for a test electron placed at the rear of the bubble.  The initial parameters (at $t$=$0$) are, $n_0=1.4\times10^{17}$ cm$^{-3}$, $r_{\perp}(0)$=$0.1 /k_p$ ($\sim$$1.3\,\mu$m), $p_{||}(0)$=$\gamma_g$, $p_{\perp}(0)$=$p_{||}(0)\times\Delta \theta_0$, where $\Delta\theta_0$=$1$ mrad.  The results are presented in Fig.(\ref{fig:lwfastage}) as trajectory (a),  energy gain (b), radiation loss (c), and QED parameter (d), in up to five plasma stages. The energy gain for each stage is close to 30 GeV. The radiation loss over the five stage is nearly 90 MeV. To investigate further, energy gain with respect to various stages (\ref{fig:lwfastage}e) and radiation loss as the function of the stage number and initial transverse positions (\ref{fig:lwfastage}f) are shown. The radiation loss is nearly $5\%$ of the TeV energy gain over the entire plasma cells comprising of $38$ stages. 

\begin{figure}[ht]
     \scalebox{0.46}
     {
          \includegraphics{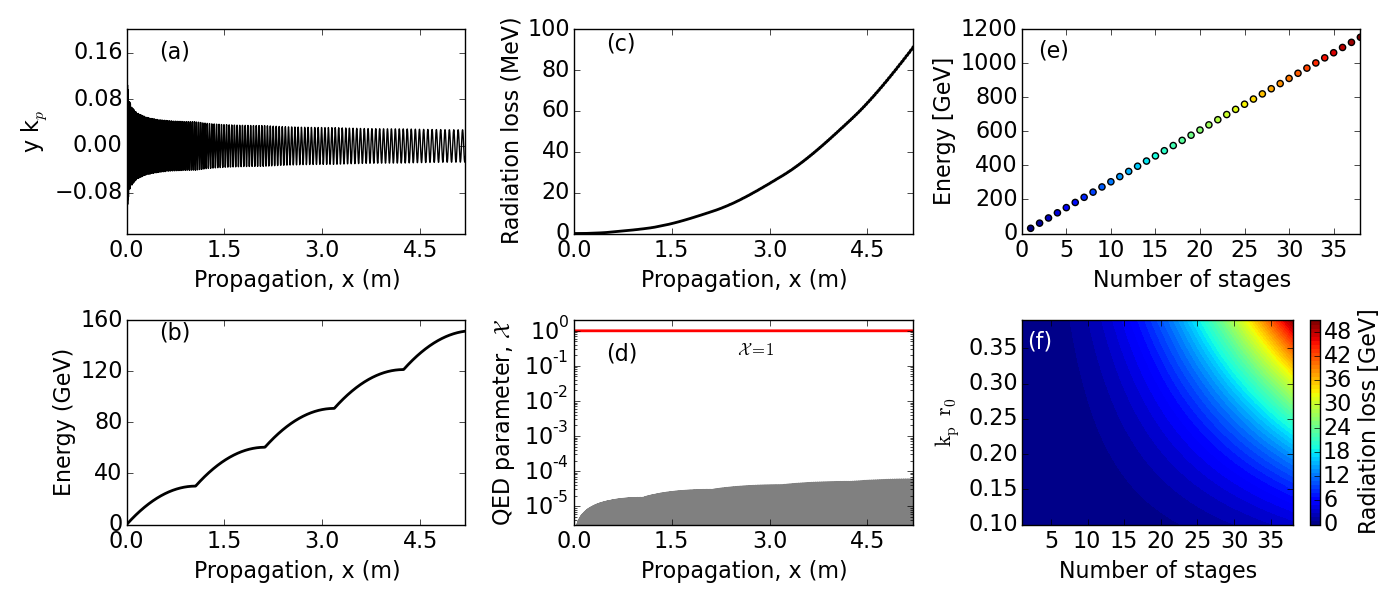} 
      }
 \caption{A test electron accelerated in a multi-staged laser-driven plasma. The ambient plasma density is $1.4\times10^{17}$cm$^{-3}$ and the corresponding plasma skin depth is $1/k_p$=$12.6\,\mu$m. For figures (a-d) the number of plasma stages is five. The initial parameters are: Transverse offset $r_{\perp}(0)$=$0.1/k_p$$(\sim1.3\,\mu m)$, $p_{||}(0)$=$\gamma_g$, $p_{\perp}(0)$=$p_{||}(0)\times\Delta \theta_0$, where $\Delta\theta_0$=$1$mrad. The results are, oscillation amplitude (a), energy gain (b), radiation loss (c), and the QED parameter (d). Figures (e-f) depict the relation between the energy gain and the stage number, and the radiation loss with respect to the initial offset and the stage number.}
    \label{fig:lwfastage}
\end{figure} 
  \section{Proton-driven Plasma Wakefield Acceleration}
  \label{sec:protondriv}
Hundreds of GeV or even TeV proton beams can potentially drive plasma wakefields over longer distances. However, currently available proton bunches are quite long. For instance, the SPS beam is approximately 12 cm, which is 100 times larger than the plasma wavelength associated with the plasma density of $4\times 10^{14}$ cm$^{-3}$; therefore it hardly excites plasma wakefields. Thanks to self-modulation instability (SMI) \cite{kupu10}, this lengthy proton beam can be transversely modulated and chopped into a train of equidistant micro bunches with the distance following the plasma wavelength. As a result, these short bunches can resonantly drive stronger plasma waves. 
 \subsection{SPS Proton-Driven Accelerator at CERN}
 After various theoretical investigations, an experiment so-called AWAKE (Advanced Wakefield Experiment at CERN) has been proposed to exploit existing SPS proton beam at CERN \cite{guoxing12} for the plasma wakefield study, which is a proof-of-principle experiment aiming to attain energy frontier charged particles \cite{Gsch16, caldwell13, Muggli15}. Theories predict that when a 400 GeV proton beam propagates into the plasma, it can excite wakefields with the strength of $\approx 1$ GV/m \cite{lotov07}. 
\begin{figure}[ht]
     \scalebox{0.5}
     {
          \includegraphics{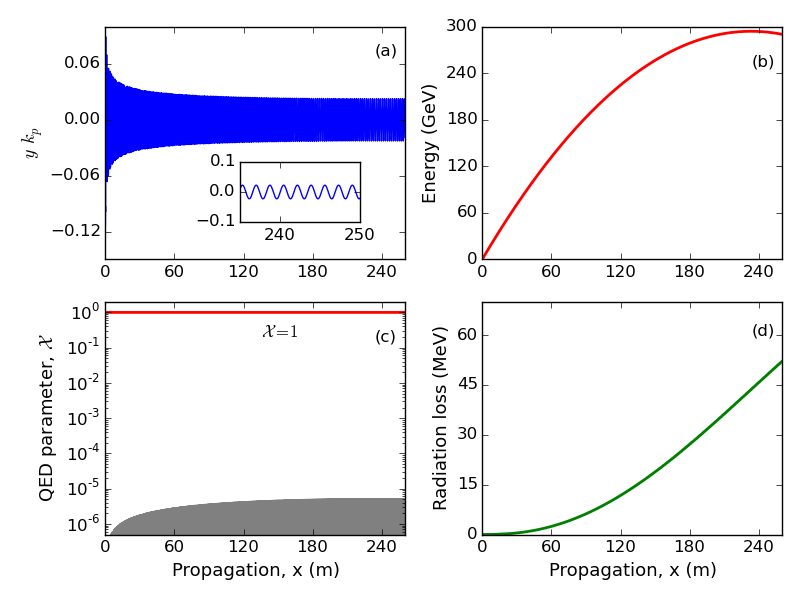}      }
 \caption{A test particle is accelerated in a proton-driven plasma wave. The plasma density is $4\times10^{14}$cm$^{-3}$, the corresponding plasma skin depth is $1/k_p\sim 254\,\mu$m, and the energy of the proton driver is $400$ GeV. The initial parameters of the test particle are $r_{\perp}(0)$=$0.02/k_p$ $(\sim5\,\mu m)$, $p_{||}(0)$=$2000$, $p_{\perp}(0)$=$p_{||}(0)\times\Delta \theta_0$, where $\Delta\theta_0$=$1$ mrad. The results are oscillation amplitude (a), energy gain (b), QED parameter (c), and the corresponding radiation loss (d).}
    \label{fig:sps}
\end{figure} 
  \begin{figure*}[t]
      \scalebox{0.3}
      {
       \includegraphics{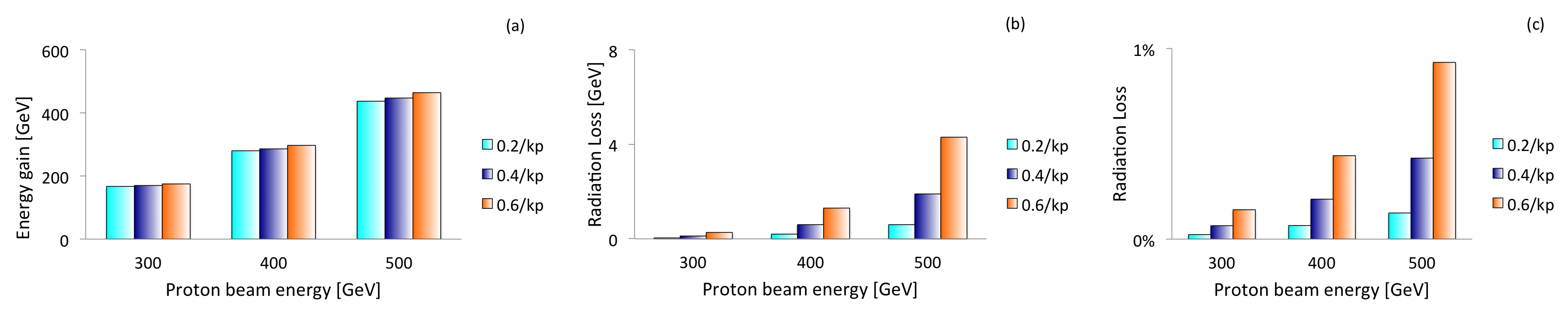} 
            }
     \caption{A test particle is accelerated in a plasma wave driven by various energetic proton beams (300 GeV, 400 GeV, 500 GeV). The values of initial transverse positions are: $0.2/k_p$ ($\sim$$50\,\mu$m),  $0.4/k_p$ ($\sim$$100\,\mu$m), and $0.6/k_p$ ($\sim$$150\,\mu$m), respectively. The energy gains (a), corresponding radiation losses (b), and loss percentages (c) are given. }
      \label{fig:spsdiff}
\end{figure*}
In the linear regime, the phase velocity of the plasma wave is smaller in comparison with the proton beam but becomes comparable when it comes to the nonlinear regime \cite{puku11}.  Considering the nonlinear bubble, we can solve the Eqs.(\ref{eqn:eqn2}) to estimate the evolution of the energy gain and the corresponding radiation loss for a test electron with an initial energy of 1 GeV (the witness electron beam is relativistic so as to catch the wakefield driven by proton bunch). The parameters for the test electron are, $n_0$=$4\times10^{14}$ cm$^{-3}$, $r_{\perp}(0)$=$0.02 /k_p$ ($\sim$$5\,\mu m)$, $p_{||}(0)$=$2000$, $p_{\perp}(0)$=$p_{||}(0)\times\Delta \theta_0$, where $\Delta\theta_0$=$1$ mrad. The test electron initially placed at the rear of the bubble. The results are shown in Fig.(\ref{fig:sps}). Figure (\ref{fig:sps}a) shows the trajectory of the test electron along with a partially magnified inset to elucidate its oscillatory nature. Figure~({\ref{fig:sps}b) shows the corresponding energy gain. Figure~(\ref{fig:sps}c) illustrates the QED parameter, which is negligible. It thus confirms the inclusion of the radiation reaction force in the equations of motion is reasonable. Fig.(\ref{fig:sps}d) shows the radiation loss, which is less than $0.02\%$ of the energy gain.

According to Kustyukov {\it et al.} \cite{kone06}, the radiation loss of an electron per unit distance can be estimated as $1.5\times10^{-45}\lbrace n_0)[cm^{-3}]\, r_0[\mu m] \, \gamma\rbrace^2$ MeV/cm, where   $r_0$ is the oscillation amplitude, and $\gamma$ is the energy gain of an electron.  
Substituting the corresponding values in the case of Fig.(\ref{fig:sps}), it gives the radiation loss of nearly 48 MeV over a 240 m long plasma, which is comparable to the numerical value ({\it ref.} Fig.~(\ref{fig:sps}d)).
Initially different transverse offsets ($r(0) k_p$=0.2, 0.4, 0.6) of the test electron are calculated for cases of various proton beam Lorentz factors ($\gamma_{pr}$=300, 400, 500), and the results are depicted in Fig.~(\ref{fig:spsdiff}). It indicates that the larger the initial offset, the greater the damping force acting on the test electron. For the close to on-axis electrons, i.e., when $r_0 \ll c/\omega_p$ for a plasma density of $4\times10^{14}$ cm$^{-3}$, the radiation reaction is negligible compared to the energy gain ({\it ref.} Fig. \ref{fig:sps}d). However, for off-axis electrons, such radiation reaction force is observable, as shown in Fig.~(\ref{fig:spsdiff}c). Also, the off-axis electrons degrade the entire beam quality and are not suitable for applications.
\subsection{LHC Beam-Driven $e^{-}e^{+}$ Linear Collider}
 A recently proposed multi-TeV $e^{-}e^+$ linear collider based on the same plasma wakefield mechanism, but the energy of the proton driver is much higher than the SPS proton beam. According to the theoretical calculation \cite{Guoxing14}, the collider designed has a centre-of-mass energy of 2 TeV, and $e^+$  and $e^-$ beams carry an energy of 1 TeV each. Theoretically, a TeV-range proton driver, driving in an ambient plasma density of $10^{14}$cm$^{-3}$, is a necessity to attain a TeV-range witness beam. Also, two drivers are required for accelerating $e^+$ and $e^-$ separately. The setup of this TeV collider can conceptualise as follows. Two proton beams, extracted from the LHC beamlines, are located at the opposite ends of a straight tunnel \cite{Guoxing14}. They excite plasma waves in two sufficiently long plasma cells for the high energy acceleration of the witness beams ($e^{-}$ and $e^+$). The accelerated beams are then transported and focused at the interaction point, which is located in the centre of the assigned straight line. 
   \begin{figure}[h]
     \scalebox{0.5}
     {
          \includegraphics{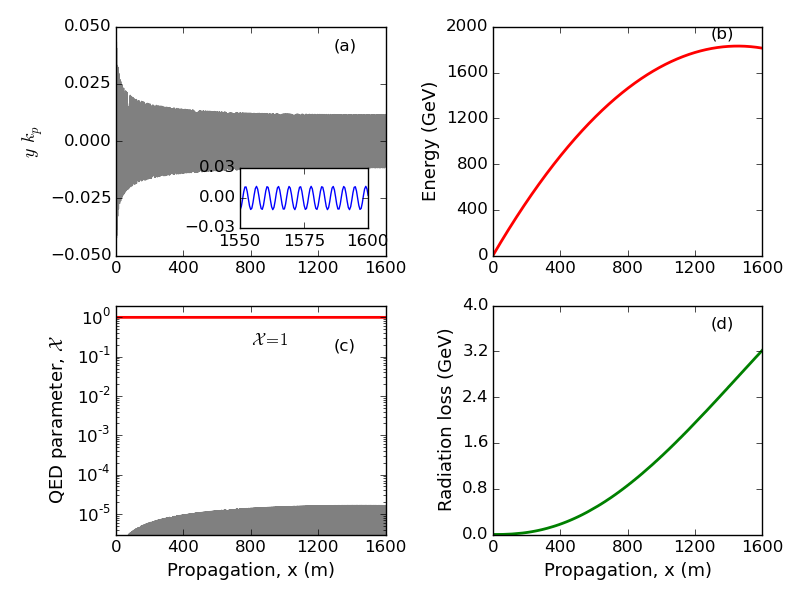}
      }
      \caption{A test particle is accelerated in a proton-driven plasma wave. The ambient plasma density is $4\times10^{14}$cm$^{-3}$ and the energy of the proton driver is 1 TeV (LHC). The initial parameters of the test particle are $r_{\perp}(0) = 0.02 /k_p$ $(\sim5\,\mu m)$, $p_{||}(0)=2000$, $p_{\perp}(0)=p_{||}(0)\times\Delta \theta_0$, where $\Delta\theta_0 =$1 mrad. The results shown are oscillation amplitude (a), energy gain (b), QED parameter (c), and the corresponding radiation loss (d).}
    \label{fig:lhc}
\end{figure} 
  \begin{figure*}[t]
      \scalebox{0.31}
      {
        \includegraphics{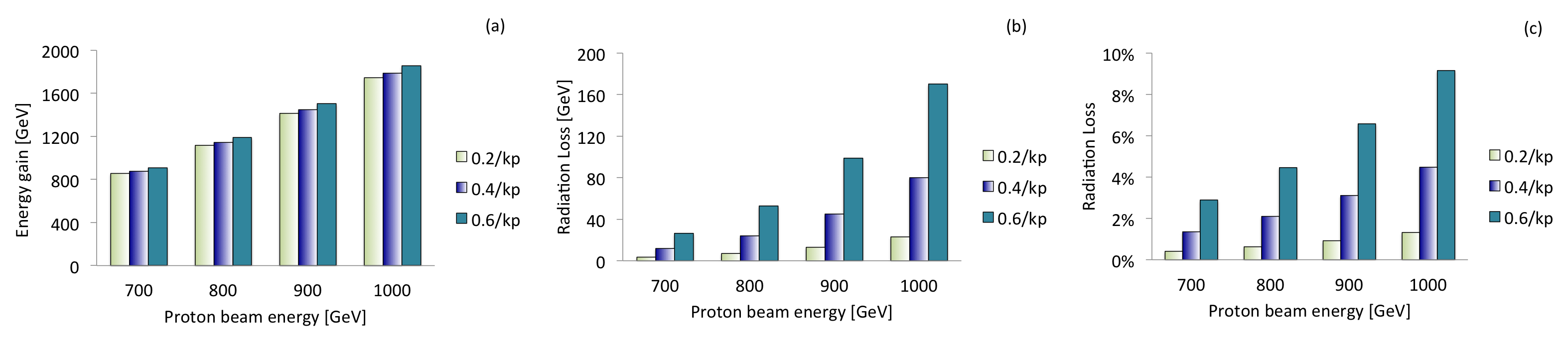} 
            }
    \caption{ A test particle is accelerated in a plasma wave driven by various energetic proton beams ($700 \dots 1000$ GeV). The values of initially different  transverse positions are: $0.2/k_p$ (=$50\,\mu$m),  $0.4/k_p$ (=$100\,\mu$m), and $0.6/k_p$ (=$150\,\mu$m). The energy gains (a), corresponding radiation losses (b), and the loss percentages (c) are given. }
    \label{fig:various}
\end{figure*}

 In a proton-driven plasma accelerator, a 1 TeV driver results in a dephasing length of nearly 1.5 km in an ambient plasma density of $4\times 10^{14}$cm$^{-3}$. Given a field gradient of 1 GV/m, the energy gain of a witness beam would be around 1.5 TeV. Now the question is, what is the damping loss before the collision of beams? The amount of radiation loss can be estimated by the numerical calculation of Eqs.~\ref{eqn:eqn2}(a-d), similar to the case with a lower energy proton beam. The initial parameters are:  $n_0= 4\times10^{14}$ cm$^{-3}$, $r_{\perp}(0) = 0.02 /k_p$ ($\sim$$5\,\mu$m), $p_{||}(0)=2000$, $p_{\perp}(0)=p_{||}(0)\times\Delta \theta_0$, where $\Delta\theta_0 =1mrad$.  The results are shown in Fig.(\ref{fig:lhc}). According to Kustyukov {\it et al.} \cite{kone06}, the following values, $r_0$=$2.5\,\mu$m, $1800$ GeV, and $n_0=4\times 10^{14}$ cm$^{-3}$, the radiation loss is nearly 3 GeV for a 1600 m long plasma. This is comparable to the numerical value ({\it ref.} Fig.~(\ref{fig:lhc}d)). The radiation loss is nearly $0.2\%$ of the energy gain.
 
 To understand the dependence of the radiation damping on the amplitude of transverse oscillations, initially different transverse positions ($r(0) k_p$=$0.2\,, 0.4\,, 0.6$) are used. Proton beams with different energies ($700 \dots 1000$ GeV) are used to drive the plasma waves. The results are shown in Fig.~(\ref{fig:various}). It implies that the larger the initial transverse position, the greater the damping loss of the test electron. For the close to on-axis electrons, i.e., when $r_0 \ll c/\omega_p$ for a plasma density of $4\times10^{14}$ cm$^{-3}$, the radiation loss is essentially small compared to the energy gain ({\it ref.} Fig.~\ref{fig:lhc}d). However, for the off-axis electrons, the radiation reaction loss is relatively appreciable (see Fig.~\ref{fig:various}c). 
 
\section{Conclusions}
\label{sec:conclusion}
In summary, we have numerically examined the radiation reaction of a charged particle accelerated by a laser-driven multi-stage plasma wakefield accelerator and proton-driven single-stage plasma wakefield accelerator. We have found that the radiation loss is not negligible in both cases. In the case of multi-stage laser-driven plasma accelerator, the radiation loss is close to 5\% of the TeV energy gain for electron close to the axis with an initial transverse offset $r_0\le 10\,\mu$m. In the case of the SPS proton beam as a driver, the radiation reaction is negligible for close to on-axial electrons accelerated up to the dephasing length. However, for the $e^{-}e^{+}$ collider case, the radiation reaction is appreciable even if electrons are injected close to on-axis. When the initial transverse displacement of the charged particle is higher but less than the plasma skin depth, the loss due to the radiation reaction increases. This loss of the accelerated electron can be up to 10\% for the collider case with an initial transverse offset $r_0 \ge 150 \,\mu$m. For the off-axis electrons, apart from substantial radiation loss, the beam quality is also not acceptable for the future applications. Therefore, close to on-axis electron injection is preferred for any future high-energy plasma-based accelerators.
\section*{Acknowledgements}
\noindent We gratefully acknowledge the support of  the Cockcroft Institute Core Grant and the STFC AWAKE (GRANT ST/P002005/1) .


\begin{thebibliography}{99}
\bibitem{Mangles2004}
S. P. D. Mangles, C. D. Murphy, Z. Najmudin, A. G. R. Thomas, J. L. Collier, A. E. Dangor, E. J. Divall, P. S. Foster, J. G. Gallacher, C. J. Hooker, D. A. Jaroszynski, A. J. Langley, W. B. Mori, P. A. Norreys, F. S. Tsung, R. Viskup, B. R. Walton, and K. Krushelnick,
 Monoenergetic beams of relativistic electrons from intense laser-plasma interactions. {\it Nature}, {\bf 431}, 535-538 (2004).

\bibitem{Geddes2004}
 C. G. R. Geddes, Cs. Toth, J. van Tilborg, E. Esarey, C. B. Schroeder, D. Bruhwiler, C. Nieter, J. Cary, and W. P. Leemans,
High-quality electron beams from a
laser wakefield accelerator using plasma-channel guiding. {\it Nature}, {\bf 431}, 538-541 (2004).
\bibitem{Faure2004} 
J. Faure, Y. Glinec, A. Pukhov, S. Kiselev, S. Gordienko, E. Lefebvre, J.-P. Rousseau, F. Burgy, and V. Malka,
A laser-plasma accelerator producing monoenergetic electron beams.
{\it Nature}, {\bf 431}, 541-544 (2004).
\bibitem{lena+06} 
W. P. Leemans, B. Nagler, A. J. Gonsalves, Cs. T$\grave{o}$th, K. Nakamura, C. G. R. Geddes, E. Esarey, C. B. Schroeder, and S. M. Hooker,
GeV electron beams from a centimetre-scale accelerator.
{\it Nature Physics}, {\bf 2}, 696-699 (2006).
\bibitem{Islam15}
M. R. Islam, E. Brunetti, R. P. Shanks, B. Ersfeld, R. C. Issac, S. Cipiccia,  M. P. Anania, G. H. Welsh, S. M. Wiggins, A. Noble,  R. A. Cairns, G. Raj, and D. A. Jaroszynski, Near-threshold electron injection in the laser-plasma wakefield accelerator leading to femtosecond bunches. {\it New Journal of Physics}, {\bf 17}, 093033 (2015).
\bibitem{Brunetti10}
E. Brunetti, R. P. Shanks, G. G. Manahan, M. R. Islam, B. Ersfeld, M. P. Anania, S. Cipiccia, R. C. Issac, G. Raj, G. Vieux, G. H. Welsh,  S. M. Wiggins, and D. A. Jaroszynski, Low emittance, high brilliance relativistic electron beams from a laser-plasma accelerator. {\it Phys. Rev. Lett.}, {\bf 105}, 215007 (2010).
\bibitem{clayton+10} 
C. E. Clayton, J. E. Ralph, F. Albert, R. A. Fonseca, S. H. Glenzer, C. Joshi, W. Lu, K. A. Marsh, S. F. Martins, W. B. Mori, A. Pak, F. S. Tsung, B. B. Pollock, J. S. Ross, L. O. Silva, and  D.H. Froula,
Self-guided laser wakefield acceleration beyond 1 GeV using ionization-induced injection. {\it Phys. Rev. Lett.}, {\bf 105}, 105003 (2010).
\bibitem{pume+02} 
A. Pukhov and J. Meyer-ter-Vehn,  Laser wake field acceleration: the highly non-linear broken-wave regime. {\it Appl. Phys. B}, {\bf 74}, 355-361 (2002).
\bibitem{Esarey2} 
E. Esarey, C. B. Schroeder, and W. P. Leemans, Physics of laser-driven plasma-based electron accelerators {\it Rev. Mod. Phys.},{\bf 81}, 1229 (2009).
\bibitem{Leemans2} 
W. P. Leemans, A. J. Gonsalves, H. S. Mao, K. Nakamura, C. Benedetti, C. B. Schroeder,
C. Toth, J. Daniels, D. E. Mittelberger, S. S. Bulanov, J. L. Vay, C. G. Geddes, and
E. Esarey, Multi-GeV electron beams from capillary-discharge-guided subpetawatt laser pulses in the self-trapping Regime, 
{\it Phys. Rev. Lett.},{\bf 113}, 245002 (2014).
\bibitem {stti+16}
S. Steinke, J. van Tilborg, C. Benedetti, C. G. R. Geddes, C. B. Schroeder, J. Daniels, K. K. Swanson, A. J. Gonsalves, K. Nakamura, N. H. Matlis, B. H. Shaw, E. Esarey, W. P. Leemans, Multistage coupling of independent laser-plasma accelerators, {\it Nature}, {\bf 530}, 190-193(2016).
\bibitem{Leemans3} 
W. Leemans and E. Esarey, Laser-driven plasma-wave electron accelerators. {\it Phys. Today}, {\bf 62 (3)}, 44 (2009).
\bibitem{kat+86}
T. Katsouleas, Physical mechanisms in the plasma wake-field accelerator. {\it Phys. Rev A}, {\bf 33}, 2056 (1986).
\bibitem{puko+08}
A. Pukhov and I. Kostyukov, Control of laser-wakefield acceleration by the plasma-density profile. {\it Phys. Rev. E}, {\bf 77}, 025401(R) (2008).
\bibitem{eskr+94}
E. Esarey, J. Krall, and P. Sprangle, Envelope analysis of intense laser pulse self-modulation in plasmas. {\it Phys. Rev. Lett.}, {\bf 72}, 2887 (1994).

\bibitem{ab+92}
L. A. Abramyan, A. G. Litvak, V. A. Mironov, and A. M. Sergeev,
Self-focusing and relativistic waveguiding of an ultrashort laser pulse in a plasma.  {\it Sov. Phys. JETP}, {\bf 75}, 978 (1992).
\bibitem{yanovsky08}
V. Yanovsky, V. Chvykov, G. Kalinchenko, and K. Krushelnick, Ultra-high intensity 300-TW laser at 0.1Hz repetition rate. {\it Opt. Express}, {\bf 16}, 2109-14 (2008).
\bibitem{eli}
T. Tajima, B. Barish, C. Barty, S. Bulanov, P. Chen, J. Feldhaus, J. Hajdu, C. Keitel, J. Kieffer, D. Ko, W. Leemans, D. Normand, L. Palumbo, K. Rzazewski, A. Sergeev, Z. Sheng, F. Takasaki, and M. Teshima, Science of Extreme Light Infrastructure. {\it  AIP Proceedings of LEI Conference 1228 -Light at Extreme Intensities "Opportunities and Technological Issues of the Extreme Light Infrastructure''}, Ed. D. Dumitras (AIP, NY, 2010) p.11.
\bibitem{jackson99}
J. D. Jackson, Classical electrodynamics. {\it Wiley}, (1999).
\bibitem{hu16}
Xing-Long Zhu, Tong-Pu Yu, Zheng-Ming Sheng, Yan Yin, Ion Cristian Edmond Turcu, and Alexander Pukhov, Dense GeV electron-positron pairs generated by lasers in near-critical-density plasmas. {\it Nature Communications}, {\bf 7}, 13686 (2016).
\bibitem{sch51}
Julian Schwinger, On gauge invariance and vacuum polarization. {\it Phys. Rev.} {\bf 82} 664 (1951).
\bibitem{simo+11} 
S. Cipiccia, M. R. Islam, B. Ersfeld, R. P. Shanks, E. Brunetti, G. Vieux, X. Yang, R. C. Issac, S. M. Wiggins, G. H. Welsh,	M. P. Anania, D. Maneuski,	 R. Montgomery, G. Smith, M. Hoek,	D. J. Hamilton,	N. R. C. Lemos, D. Symes, P. P. Rajeev, V. O. Shea,	J. M. Dias, and D. A. Jaroszynski, Gamma-rays from harmonically resonant betatron oscillations in a plasma wake. {\it Nature Physics}, {\bf 7}, 867-871 (2011).
\bibitem{blcl+07}
I. Blumenfeld, C. E. Clayton, Franz-Josef Decker, M. J. Hogan, C. Huang, R. Ischebeck, R. Iverson, C. Joshi, T. Katsouleas, N. Kirby, W. Lu, K. A. Marsh, W. B. Mori, P. Muggli, E. Oz, R. H. Siemann, D. Walz, and M. Zhou, Energy doubling of 42 GeV electrons in a metre-scale plasma wakefield accelerator.  {\it Nature}, {\bf 445}, 741-744 (2007).
\bibitem{Litos}
M. Litos, E. Adli, W. An, C.I. Clarke, C.E. Clayton, S. Corde, J.P. Delahaye, R.J. England, A.S. Fisher, J. Frederico, S. Gessner, S.Z. Green, M.J. Hogan, C. Joshi, W. Lu, K.A. Marsh, W.B. Mori, P. Muggli, N. Vafaei-Najafabadi, D. Walz, G. White, Z. Wu, V. Yakimenko, G. Yocky, High-efficiency acceleration of an electron beamin a plasma wakefield accelerator. {\it Nature}, {\bf 515},v92 (2014).
\bibitem{Corde}
S. Corde, E. Adli, J.M. Allen, W. An, C.I. Clarke, C.E. Clayton, J.P. Delahaye, J. Frederico, S. Gessner, S.Z. Green, M.J. Hogan, C. Joshi, N. Lipkowitz, M. Litos, W. Lu, K.A. Marsh, W.B. Mori, M. Schmeltz, N. Vafaei-Najafabadi, D. Walz, V. Yakimenko, G. Yocky, Multi-gigaelectronvolt acceleration of positrons in a self-loaded plasma wakefield. {\it Nature}, {\bf 524}, 442 (2015).
\bibitem{Gessner} 
S. Gessner, E. Adli, J. M. Allen, W. An, C. I. Clarke, C. E. Clayton, S. Corde, J. P. Delahaye, J. Frederico, S. Z. Green, C. Hast, M. J. Hogan, C. Joshi, C. A. Lindstroem, N. Lipkowitz, M. Litos, W. Lu, K. A. Marsh, W. B. Mori, B. O'Shea, N. Vafaei-Najafabadi, D. Walz, V. Yakimenko, and G. Yocky, Demonstration of a positron beam-driven hollow channel plasma wakefield accelerator. {\it Nature Communications}, {\bf 7}, 11785 (2016).
\bibitem{Seryi} 
A. Seryi, M. Hogan, S. Pei, T. Raubenheimer, P.Tenenbaum, T. Katsouleas, C. Huang, C. Joshi, W. Mori, P. Muggli, A concept of plasma wake field acceleration linear collider (PWFA-LC). SLAC-PUB-13766 (2009).
\bibitem{Adli} 
E. Adli, J. P. Delahaye, S. J. Gessner, M. J. Hogan, T. Raubenheimer, W. An, C. Joshi, W. Mori, A beam driven plasma-wakefield linear collider: from Higgs factory to multi-TeV. SLAC-PUB-15426 (2013).
\bibitem{caldwell13}
A. Caldwell, E. Gschwendtner, K. Lotov, P. Muggli, and M. Wing, AWAKE Design Report, A Proton-Driven Plasma
Wakefield Acceleration Experiment at CERN. {\it Internal Note CERNSPSC-2013-013, CERN}, Geneva, Switzerland (2013).
\bibitem{Guoxing14}
G. Xia, O.Mete, A.Aimidula, C.P.Welsch, S.Chattopadhyay, S. Mandry, and M.Wing, Collider design issues based on proton-driven plasma
wakefield acceleration. {\it  Nuclear Instruments and Methods in Physics Research A}, {\bf 740}, 173-179, (2014).

\bibitem{Piazza12}
A. Di Piazza, C. M\"{u}ller, K. Z. Hatsagortsyan, and C. H. Keitel, Extremely high-intensity laser interactions with fundamental quantum systems. {\it Rev. Mod. Phys.}, {\bf 84} 1177 (2012). 
\bibitem{lali80}
L. D.  Landau and E. M. Lifshitz, The classical  Theory of Fields. {\bf 2}, {\it Oxford: Butterworth-Heinemann} (1980). 
\bibitem{essp96}
E. Esarey, P. Sprangle, J. Krall, and A. Ting,  Overview of plasma-based accelerator
concepts. {\it Plasma Science, IEEE Transactions}, {\bf 24}, 252 288, (1996).
\bibitem{rocl88}
J. B. Rosenzweig, D. B. Cline, B. Cole, H. Figueroa, W. Gai, R. Konecny, J. Norem,
P. Schoessow, and J. Simpson, Experimental observation of plasma wakefield acceleration. {\it Phys. Rev. Lett.}, {\bf 61}, 98, (1988). 
\bibitem{luhu+06}
W. Lu, C. Huang, M. Zhou, and M. Tzoufras, A nonlinear theory for multidimensional relativistic plasma wave wakefields, {\it Phys. of Plasmas}, {\bf 13}, 056709 (2006).
\bibitem{kopu+04}
I.  Kostyukov, A. Pukhov, and S.  Kiselev,  Phenomenological theory of laser-plasma interaction in ``bubble" regime.  {\it Phys. of Plasmas}, {\bf 11}, 5256 (2004).
\bibitem{lees+09}
W. Leeman and W. Esarey, Laser-driven plasma-wave electron accelerators, {\it Physics Today}, {\bf 62}, 3,44 (2009).

\bibitem{kupu10}
N. Kumar, A. Pukhov, and K. Lotov, Self-Modulation Instability of a Long Proton Bunch in Plasmas. {\it Phys. Rev. Lett.}, {\bf 104}, 255003 (2010).
\bibitem{guoxing12}
G. Xia, R. Assmann, R. A. Fonseca, C. Huang, W. Mori, L. O. Silva, J. Vieira, F. Zimmermann, and P. Muggli, A proposed demonstration of an experiment of proton-driven plasma wakefield acceleration based on CERN SPS. {\it Journal of Plasmas Physics}, {\bf 78}, 347 (2012).
\bibitem{Gsch16}
 E. Gschwendtner, E. Adli, L. Amorim, R. Apsimon, R.Assmann, A.-M. Bachmann, F. Batsch, J. Bauche, 
 V. K. Berglyd Olsen, M. Bernardini, R. Bingham, B. Biskup, T. Bohl, C. Bracco, P. N. Burrows, G. Burt, {\it et al.}, AWAKE, The Advanced Proton Driven Plasma Wakefield Acceleration Experiment at CERN. {\it Nuclear Instruments and Methods in Physics Research Section A: Accelerators, Spectrometers, Detectors and Associated Equipment}, {\bf 829}, 76-82 (2016).

\bibitem{Muggli15}
P. Muggli, AWAKE, proton-driven plasma wakefield experiment at
CERN. {\it IPAC15}, Richmond, USA (2015).
\bibitem{lotov07}
K. V. Lotov, Acceleration of positrons by electron beam-driven wakefelds in a plasma. {\it Physics of Plasmas}, {\bf 14}, 023101, (2007).
\bibitem{puku11}
A. Pukhov, N. Kumar, T. T\"{u}ckmantel, A. Upadhyay, K. Lotov, P. Muggli, V. Khudik, C. Siemon, and G. Shvets, Phase Velocity and Particle Injection in a Self-Modulated Proton-Driven Plasma Wakefield Accelerator. {\it Phys. Rev. Lett.}, {\bf 107}, 145003 (2011).
\bibitem{kone06}
I. Yu. Kostyukov, E.N. Nerush, and A. Pukhov, Synchrotron radiation losses in laser-plasma accelerators.
{\it Problems of atomic science and technology. No 2. Series: Nuclear Physics Investigations}, {\bf 46}, 169-171 (2006).

\end{thebibliography}
\end{document}